\documentstyle[12pt]{article}
\newcommand{\doublespace}{\renewcommand{\baselinestretch}{1.75}
   \Large\normalsize}

\renewcommand{\ref}[1]{\raisebox{.6ex}{[#1]}}

\newcommand{\be}{\begin{equation}}
\newcommand{\ee}{\end{equation}}

\newcommand{\ol}{\overline }

\begin{document}

\doublespace

\title{ Magnetic Field Effect on an Electron Tunneling
 out of a Confining Plane }

\author{Ping Ao     \\
Department of Physics, FM-15                 \\
University of Washington, Seattle, WA 98195  \\ }

\maketitle

\begin{abstract}
The influence of a magnetic field on the tunneling of an electron out of a
confining plane is studied by a path integral method. We map this 3-d problem
on
to a 1-d one, and find that the tunneling is strongly affected by the field.
Without a perpendicular field the tunneling at zero temperature can be
completely suppressed by a large parallel field, but in the small parallel
field and low temperature limit the tunneling rate is finite. An explicit
formula is obtained in this case. A quantitative explanation without fitting
parameter to a recent experiment is provided.
\end{abstract}

\noindent
PACS${\#}$s: 73.40.Gk; 73.20.Dx;

\newpage

The tunneling of charged particles in the presence of a magnetic field has
many peculiar properties. It occurs in the study of
the escape of trapped electrons from a helium-vacuum
interface\cite{exp,ao1,menna}, the focus of the present paper.
It is, however, a general phenomenon and is widely encountered
in mesoscopic systems\cite{mesoscopic},
in semiconductor heterostructures\cite{semiconductor,breakdown},
in disorder media\cite{random}, and in Quantum Hall Systems such as the
tunneling between double layers\cite{double} and the quasiparticle
tunneling\cite{quasiparticle}.
The helium-vacuum-interface system is relatively simple, but
it shares many features of the systems in
Refs.\cite{mesoscopic,semiconductor,breakdown,random,double,quasiparticle},
such as the influences of magnitude and direction of a magnetic field and of
many-body correlations on the tunneling.
This system provides us a well controlled test ground for our understanding of
tunneling in higher dimensions.
There is a considerable experimental study of the escape rate on
it\cite{exp,menna}.
In the case of no magnetic field the tunneling process in this system is
adequately understood, which is effectively a 1-d problem even after the
consideration of many-body correlations\cite{exp,ao1}. However,
the presence of a magnetic field requires a full 3-d study of the tunneling
process. It gives rise to new effects about the magnetic
field and temperature dependence of the tunneling rate, as shown in a recent
experiment which requires a clear and detailed theoretical
explanation\cite{menna}. The purpose of the present paper
is to give a formulation of the problem in this system and to present
a detailed analytical study.
A quantitative agreement without fitting parameter with the experiment of
Ref.\cite{menna} is obtained.
Our results, as well as our path integral method which keeps
the relevant degrees of freedom but eliminates irrelevant ones by integration,
may have a wider applications in the systems of
Refs.\cite{mesoscopic,semiconductor,breakdown,random,double,quasiparticle}.
In fact, our Hamiltonian is identical to those, for example, in the study of
magnetotransportation in semiconductor heterostructures\cite{semiconductor}.

Our main results are as follows. We consider a situation in 3-d that
in one direction an electron has a metastable state,
and there is no force in the remaining two directions.
The tunneling out the metastable state is studied as a function of  both
magnitude and direction of an externally applied constant magnetic field,
and of temperature. We map this problem onto that of quantum dissipative
tunneling. The in-plane motion acts as an environment to the tunneling in
the perpendicular direction because of the parallel magnetic field.
This environmental influence is found to be subohmic when
the perpendicular component of the magnetic field $B_{\perp}=0$,
and superohmic when $B_{\perp} \neq 0$.
{}From the obtained effective action we find that
the tunneling is strongly influenced by the magnetic field. In particular,
if $B_{\perp}=0$ and the magnetic field $B_{\parallel}$ parallel to
the force free
plane is large enough there is a complete suppression of tunneling rate
at zero temperature, and the electron becomes dynamically localized.
However, if $B_{\parallel}$ is small, the tunneling rate out of the
metastable state is finite. Perturbatively,
the semiclassical action is
proportional to $B_{\parallel}^{2}$, when $B_{\perp} =0$.
In this case the calculation can be generalized to a finite but low
temperature, where we find that the semiclassical action decreases linearly
with temperature $T$.
The last two results are in a remarkable agreement with the experiment of
Ref.\cite{menna}.

Now we present the calculation leading to the above results.
For the convenience of calculation we shall first assume that the $x$- and
$y$-directions have  weak harmonic potentials. We take them to be zero,
that is, $\omega_{0} \rightarrow 0$, in the end of the calculation.
The Hamiltonian for an electron with mass $m$ and charge $e$ is then
\be
   H = \frac{1}{2m} [ {\bf P} - \frac{e}{c} {\bf A}({\bf R}) ]^{2} +
       \frac{1}{2} m \omega_{0}^{2} ( x^{2} + y^{2} ) + V(z) \;  ,
\ee
with the vector potential ${\bf A}$ determined by
$ \nabla\times {\bf A} = {\bf B} $.
Here the metastable point of the potential is taken at $z=0$. For large enough
$z$, $V(z) = - F z$, and there is a barrier separating this region and $z=0$.
The external magnetic field is tilted:
${\bf B} = (B_{\parallel}, 0 , B_{\perp} )$.
In accordance with the calculation of tunneling,
the vector potential will be taken as
\be
   {\bf A} =  (0, B_{\perp} x - B_{\parallel} z, 0) \;,
\ee
and it can be shown that results are independent of the choice of gauge
because of the periodic boundary condition in the calculation of the tunneling
rate. This is a 3-d tunneling problem because the magnetic field couples the
motions in all directions.

The tunneling is described by the Euclidean action\cite{caldeira,jain}
\be
   S[{\bf R}(\tau) ] = \int_{0}^{\hbar\beta } d\tau \left[ \frac{1}{2}
       m \dot{\bf R}^{2} +
       i \frac{e }{c } (B_{\perp} x - B_{\parallel} z ) \dot{y} +
          V(z) + \frac{1}{2} m \omega_{0}^{2} ( x^{2} + y^{2} ) \right]  \; ,
\ee
where $\beta = 1/k_{B}T$ is the inverse temperature.
The tunneling rate is proportional to $\exp\{ - S_{c}/\hbar \} $, where
the semiclassical action $S_{c}$ is determined by the
bounce solution of the equation $\delta S =0$, in which the periodic boundary
condition, ${\bf R}(\hbar\beta) = {\bf R}(0) $, is required.
This is so-called the method of the imaginary part of the free
energy $F = -k_{B}T \ln Z$ \cite{hanggi} calculated from the partition function
\be
   Z = \int d{\bf R}' \; D({\bf R}'; {\bf R}' ) \; ,
\ee
with the density matrix
\be
   D({\bf R}'(0) ; {\bf R}''(\hbar\beta) )
   = \int {\cal D}{\bf R} \;
     \exp\left\{- \frac{1}{\hbar}  S[{\bf R}]   \right\} \; .
\ee
This method is identical to the WKB method at zero temperature, and allows us
to have a unified treatment of the escape rate for finite temperatures.

We are interested in the electron tunneling out of the metastable state $z=0$
in the $z$-direction.
After the tunneling other degrees of freedoms, $x$ and $y$,
can take any allowed value. Therefore the summation over final states,
integrations over the $x$ and $y$ coordinates, will be taken in the
calculation of the partition function. Those are gaussian integrals.
To integrate over the $x$-coordinate, we perform a Fourier
transformation on the time interval $[0, \; \hbar\beta]$:
\be
   (x(\tau), y(\tau), z(\tau) ) = \frac{1}{\hbar\beta}
\sum_{n=-\infty}^{\infty}
             (x_{n}, y_{n}, z_{n} ) e^{i \nu_{n} \tau } \; .
\ee
Here $\nu_{n} = 2\pi n /\hbar\beta \; $.
The action $S$ of eq.(3) can be then rewritten as
\[
   S[\{ {\bf R}_{n} \} ] =
      \frac{1}{\hbar\beta} \sum_{n=-\infty}^{\infty}
      \left[ \frac{1}{2} m \nu_{n}^{2} z_{n} z_{-n} + V(z)_{n}
      + \frac{1}{2} ( m \nu_{n}^{2} + m \omega_{0}^{2}  ) x_{n} x_{-n} \right]
\]
\be
   + \frac{1}{\hbar\beta} \sum_{n=-\infty}^{\infty}
   \left[ \frac{1}{2} ( m \nu_{n}^{2} + m \omega_{0}^{2} ) y_{n} y_{-n}
     + \frac{e}{c}( B_{\perp} x_{n} - B_{\parallel} z_{n})
      \nu_{-n} y_{-n} \right] \; .
\ee
Here $\nu_{n} = - \nu_{-n}$ has been used.
Now, we can shift the origin of $\{y_{n} \}$, and integrate over them.
Then we obtain the effective action as
\[
    S_{eff}[ \{ x_{n}, z_{n} \} ] =
      \frac{1}{\hbar\beta} \sum_{n=-\infty}^{\infty}
      \left[ \frac{1}{2} m \nu_{n}^{2} z_{n} z_{-n} + V(z)_{n}
      + \frac{1}{2} ( m \nu_{n}^{2} + m \omega_{0}^{2}  ) x_{n} x_{-n} \right]
\]
\be
      - \frac{1}{\hbar\beta} \sum_{n=-\infty}^{\infty} \frac{1}{2}
         \frac{ (\frac{e}{c} )^{2} \nu_{n} \nu_{-n}          }
          { m\nu_{n}^{2} + m\omega_{0}^{2} }
        ( B_{\perp} x_{n} - B_{\parallel} z_{n} )
        ( B_{\perp} x_{-n} - B_{\parallel} z_{-n}) \; .
\ee
Similarly, the integration over $\{ x_{n} \}$, can be performed. The resulting
effective action is
\[
    S_{eff}[ z(\tau) ] = \int_{0}^{\hbar\beta } d\tau
    \left[ \frac{1}{2} m \dot{z}^{2} + V(z) \right]
    - \frac{1}{2} \int_{0}^{\hbar\beta } d\tau \int_{0}^{\hbar\beta } d\tau'
      [ \ol{g}_{1}(\tau - \tau' )
\]
\be
      - g_{2}(\tau - \tau' ) ] [ z(\tau) - z(\tau') ]^{2} \; ,
\ee
with the first kernel
\be
   \ol{g}_{1}(\tau) = \frac{1}{\hbar\beta}
     \sum_{n=-\infty}^{\infty} \frac{1}{2}
     \frac{ (\frac{e}{c} B_{\parallel} )^{2} \nu_{n}^{2} }
          { m\nu_{n}^{2} + m\omega_{0}^{2}  } \;
             e^{ i\nu_{n} \tau } \; ,
\ee
and the second kernel
\be
    g_{2}(\tau) = \frac{1}{\hbar\beta} \sum_{n=-\infty}^{\infty} \frac{1}{2}
         \frac{\left[ B_{\perp} B_{\parallel} ( \frac{e}{c} )^{2}
                 \nu_{n} \nu_{-n}  \right]^{2}         }
             {  [ m\nu_{n}^{2} + m\omega_{0}^{2} ]
          \left[ \left(\frac{e}{c} B_{\perp} \right)^{2} \nu_{n}^{2}
            + ( m\nu_{n}^{2} + m\omega_{0}^{2} )^{2} \right] } \;
             e^{ i \nu_{n} \tau } \; .
\ee
Here $\int_{0}^{\hbar\beta } d\tau \; \ol{g}_{1}(\tau) = 0$ and
$\int_{0}^{\hbar\beta } d\tau \; g_{2}(\tau) = 0$
have been used in arriving above result.

Eqs.(9-11) can be simplified. We note that
the kernel $\ol{g}_{1}$ can be separated into two parts:
\be
    \ol{g}_{1}(\tau) = \frac{1}{\hbar\beta}\frac{m}{2}
    \left( \frac{e B_{\parallel} }{mc} \right)^{2}
           \sum_{n=-\infty}^{\infty} e^{ i \nu_{n}\tau }
           - g_{1}(\tau) \; ,
\ee
with
\be
   g_{1}(\tau) = \frac{m}{2} \left( \frac{e B_{\parallel} }{mc} \right)^{2}
         \frac{\omega_{0} }{2}
         \frac{ \cosh [\omega_{0} ( \frac{\hbar\beta}{2} -\tau ) ] }
              { \sinh [ \frac{ \hbar\beta \omega_{0}  }{2} ] }  \; .
\ee
The first part of the right hand of eq.(12) is simply a periodic delta
function,
which gives no contribution to the effective action and will be dropped.
We further simplify the expressions by taking
the limit of $\omega_{0} = 0$, and find that
\be
   \ol{g}_{1}(\tau) = - g_{1}(\tau ) = - \; \frac{1}{\hbar\beta}\frac{m}{2}
                  \left( \frac{e B_{\parallel} }{mc} \right)^{2}  \; ,
\ee
and
\be
   g_{2}(\tau ) = \frac{m}{2} \left( \frac{e B_{\parallel} }{mc} \right)^{2}
         \left[ \frac{ \omega_{\perp} }{2}
         \frac{ \cosh[ \omega_{\perp} (\frac{\hbar\beta}{2} -\tau ) ] }
              { \sinh[ \frac{ \hbar \beta \omega_{\perp} }{2} ] }
            - \frac{1}{\hbar\beta} \right] \; ,
\ee
with $\omega_{\perp} = B_{\perp} e /mc \; $.
Then using eqs.(14,15,9) we obtain the effective action for our problem as
\[
    S_{eff}[ z(\tau) ] = \int_{0}^{\hbar\beta } d\tau
    \left[ \frac{1}{2} m \dot{z}^{2} + V(z) \right]
   + \frac{1}{2} \int_{0}^{\hbar\beta } d\tau \int_{0}^{\hbar\beta } d\tau'
       \frac{m}{2} \left( \frac{e B_{\parallel} }{mc} \right)^{2} \times
\]
\be
        \frac{ \omega_{\perp} }{2}
        \frac{ \cosh[ \omega_{\perp} (\frac{\hbar\beta}{2} -|\tau-\tau'| ) ] }
             { \sinh[ \frac{ \hbar \beta \omega_{\perp} }{2} ] }
         [ z(\tau) - z(\tau') ]^{2}    \; .
\ee
Thus, we have mapped the original 3-d problem onto a 1-d one.
This is because the in-plane motion, $x$- and $y$-directions,
effectively behaves as a bath for $z$-direction motion\cite{caldeira}, where
the parallel magnetic field $B_{\parallel}$ serves as a coupling coefficient.
This form of effective action is easy to handle because of the existence
of a sophisticated technique
to calculate the dissipative tunneling rate\cite{caldeira}.
In terms of the dissipative tunneling\cite{caldeira},
the environmental effect in the eq.(16) is
a superohmic damping of $s=\infty$ for $B_{\perp} \neq 0$, where
the environment consists of the in-plane cyclotron motion.
If $B_{\perp} = 0$, it is then a subohmic damping of $s=0$,
where the environment consists of in-plane plane-waves.
We should point out that no approximation is used in obtaining eq.(16).
Simply by inspection of eq.(16),
one can conclude that the effective bath due to
the in-plane motion will influence the tunneling rate out of the confining
plane by affecting the semiclassical action.
A numerical calculation of the tunneling rate will be needed to cover whole
parameter region.
In the following we take $B_{\perp} = 0$,
and focus on the peculiar subohmic damping case with $B_{\parallel} \neq 0$.

{}From eq.(16), the effective action with $B_{\perp} = 0$ is
\be
   S_{eff}[z(\tau)] = \int_{0}^{\hbar\beta } d\tau \left[ \frac{1}{2}
      m \dot{z}^{2} + V(z)  \right]
    + \frac{m}{4} \left( \frac{e B_{\parallel} } {mc } \right)^{2}
      \frac{1}{\hbar\beta}
      \int_{0}^{\hbar\beta} d\tau \int_{0}^{\hbar\beta} d\tau'
      [ z(\tau) - z(\tau') ]^{2} \; .
\ee
This equation is explicitly gauge invariant under the change $y \rightarrow
y + \; constant$, because of the periodic boundary condition of $x,y$
imposed in the tunneling calculation as pointed out above.
If there is a tunneling solution, it can be shown that
\be
   \lim_{\hbar\beta \rightarrow \infty }
     \int_{0 }^{\hbar\beta } d\tau  \; z(\tau) = \; constant
\ee
for the semiclassical solution.
This is the result of the period boundary condition that
$   \lim_{\hbar\beta \rightarrow \infty } z(\hbar\beta) = z(0) =  0$ and the
finiteness of the semiclassical action.
Therefore, the effective action at zero temperature may be written as
\be
   S_{eff}[z(\tau)] = \int_{-\infty}^{\infty } d\tau \left[ \frac{1}{2}
      m \dot{z}^{2}(\tau) + V(z(\tau))
      + \frac{m}{2} \left(\frac{e B_{\parallel}}{mc }
      \right)^{2}  z^{2}(\tau) \right] \; .
\ee
It shows that the original potential $V(z)$ is renormalized to
$ V(z) + m \; ( e B_{\parallel}/mc )^{2}  z^{2}/2$ ,
and the metastable state at $z=0$ becomes more stable.
For a large $z$, the renormalized potential is positive now,
and we have a double-well like potential\cite{bhattacharya}.
If the second local potential minimum is lower than the one at $z=0$, tunneling
out of $z=0$ is finite. However, for a large enough parallel magnetic
field $B_{\parallel}$,
the second local potential minumum will be higher than the one at $z=0$.
The tunneling rate is then zero. In this case, eq.(18) is not valid.
This analysis suggests that there is a magnetic field induced localization
transition at zero temperature.
The violation of eq.(18) may serve as the indication for this
transition, or alternatively, the comparison of the local
two potential minima of the renormalized potential in eq.(19)
may determine the critical magnetic field.
Thus we have obtained that the $s=0$ dissipative environment is marginal
for localization in tunneling from a metastable state,
compared to the $s=1$ case for the tunneling splitting\cite{loc,leggett}.

Away from the localization region,
for a small parallel magnetic field $B_{\parallel}$
the tunneling rate is finite.
In particular, for a very small parallel magnetic field
and low temperatures the semiclassical action may be evaluated perturbatively:
\[
   S_{c} = \int_{-\infty}^{\infty } d\tau \left[ \frac{1}{2}
      m \dot{z}^{2}_{c}(\tau) + V( z_{c}(\tau) )  \right]
    + \frac{m}{2} \left( \frac{e B_{\parallel} }{mc} \right)^{2}
      \left[ \int_{-\infty}^{\infty} d\tau  z_{c}^{2}(\tau) \right.
\]
\be
    \left.
    - \frac{ k_{B}T }{\hbar} \left( \int_{-\infty}^{\infty}
                                 d\tau  z_{c}(\tau) \right)^{2} \right] \; ,
\ee
where $z_{c}(\tau)$ is the bounce solution at zero temperature without the
magnetic field, determined by the equation $m\dot{z}^{2}_{c}/2 = V(z_{c})$.
Eq.(20) shows a pronounced $B_{\parallel}^{2}$ dependence and
linear temperature dependence.

Now we briefly discuss the effect of finite perpendicular magnetic field
$B_{\perp}$.
For $B_{\perp} \neq 0$ the effective bath changes from $s=0$ at
$B_{\perp}=0$ to $s=\infty$. From Refs.\cite{leggett,hanggi}
we expect that the tunneling can occur
at any value of $B_{\parallel}$ and there is no localization.
Therefore by tilting the direction of the magnetic field there should be a
large change in the tunneling rate.

To conclude, we discsuss the experimental verification of the above results.
The tunneling rate of trapped electrons escaping from a helium-vacuum interface
was measured in Ref.\cite{menna} as a function of $B_{\parallel}^{2}$ and $T$
with $B_{\perp} =0$.
A numerical estimation shows the data are in the low parallel magnetic field
and low temperature limit. The comparison between the calculation according to
eq.(20) and the data of Ref.\cite{menna} is shown in the Figure.
A quantitative agreement is found in the low temperature limit.
Note that there is no fitting parameter.
The deviation in the high temperature end in the Figure
suggests that the thermal activation starts to play a role.
It is interesting to note that the corresponding critical magnetic field
according to the renormalized potential in eq.(19) is about 1.3 Tesla
for that experiment, which seems easy to realize experimentally.
In Ref.\cite{menna} it is also found that there is
no dependence of tunneling rate on $B_{\perp}$ when $B_{\parallel} = 0$.
This is implied in eq.(16), too, where by taking $B_{\parallel} = 0$ the
in-plane motion is decoupled from the tunneling in $z$-direction.
To further assure oneself of the  agreement between eq.(20)
and the experimental data, the absence of the effect of the other dynamical
correlations in the experiment need to be addressed. We discuss this below.

It has been shown that the response of the remaining 2-d electrons to the
tunneling electron is superohmic\cite{ao1}.
The influence from ripplons is not only superohmic, also weak.
The only possible ohmic effect is from the He atom scatterings. However, its
strength is exponentially small at low temperature, and can be ignored at low
temperatures. Because the superohmic bath has a weak influence on tunneling
in the absence of a magnetic field\cite{leggett,hanggi}, the tunneling is
therefore dominated by the adiabatic potential.
Therefore the adiabatic potential is good in the explanation of the
experiments\cite{exp} and dynamical correlations from those environments
are weak\cite{ao1}.
Furthermore, it has been shown recently that, the superohmic bath has a weak
influence on the tunneling rate even in the presence of a magnetic
field\cite{ao2}.
Therefore the adiabatic potential can be used in the tunneling rate
calculation with a magnetic field, too.
Using the adiabatic potential
first proposed by Iye {\it et al} in Ref.\cite{exp}, we have calculated the
semiclassical action by eq.(20) as shown in the Figure.
In the calculation the Stark shift and the distance of the 2-d electron layer
from the helium-vacuum interface have been considered.

\noindent
{\bf Acknowledgements:}
The assistance from Yong Tan on numerical calculation is appreciated.
This work was supported by US National Science Foundation under Grant Nos
DMR-8916052 and DMR-9220733.

\newpage

Figure Caption.

Figure. The circles are the experimental data from Ref.[3].
The solid line is the calculation according to eq.(20), in which
the 2-d electron density $n= 8.7\times 10^{7}/cm^{2}$
and the external electric field $E = 30 V/cm$ are used.

\end{document}